\documentclass[aps,prb,twocolumn,floatfix,citeautoscript,nofootinbib,superscriptaddress]{revtex4-2}
\usepackage{amsbsy}
\usepackage{latexsym,epsfig,graphicx}
\usepackage{dcolumn}
\usepackage{multirow}
\usepackage{diagbox}
\usepackage{graphicx}
\usepackage{subfigure}
\usepackage{comment}
\usepackage{color}
\usepackage{xcolor}
\usepackage{soul}
\usepackage{bm}
\usepackage{mathrsfs}
\usepackage{amsfonts}
\usepackage{amsmath}
\usepackage{amssymb}
\usepackage{xspace}
\usepackage{epstopdf}
\usepackage{tabularx}
\usepackage{longtable}
\usepackage{bbding}
\usepackage[colorlinks=true, letterpaper=true, pdfstartview=FitV, linkcolor=blue, citecolor=blue, urlcolor=blue]{hyperref}
\usepackage[normalem]{ulem}

\begin{document}

\title{Symmetry-enforced third-order nonlinear thermal Hall effects in altermagnets}
\author{Yun-Mei Li}
\email{yunmeili@zju.edu.cn}
\affiliation{Center for Quantum Matter, School of Physics, Zhejiang University, Hangzhou 310027, China}
\author{Jiacheng Yao}
\affiliation{Center for Quantum Matter, School of Physics, Zhejiang University, Hangzhou 310027, China}
\author{Hua Wang}
\affiliation{Center for Quantum Matter, School of Physics, Zhejiang University, Hangzhou 310027, China}
\author{Kai Chang}
\email{kchang@zju.edu.cn}
\affiliation{Center for Quantum Matter, School of Physics, Zhejiang University, Hangzhou 310027, China}

\begin{abstract}
  Thermal Hall effect (THE) is a powerful probe of material properties, even in insulators. Here we investigate the thermal response of altermagnets by developing a theory on Berry curvature driven nonlinear THE from both electrons and magnons. We identify symmetry conditions under which the third‑order THE dominates while linear and second‑order contributions vanish. Notably, $d$-wave and $i$-wave altermagnets with out-of-plane N\'{e}el order satisfy these conditions, whereas in‑plane order yields a dominant linear THE. Using KV$_{2}$Se$_{2}$O (a $d$-wave altermagnetic metal) and MnF$_{2}$ (an altermagnetic insulator) as concrete examples, we show nonzero third‑order nonlinear THE from electrons and magnons, respectively, both exhibiting a $\pi$-periodic dependence on the direction of temperature gradient. These symmetry‑guaranteed Berry‑curvature phenomena provide a diagnostic for altermagnetism in candidate materials and enable determination of the N\'{e}el‑vector orientation. 
\end{abstract}

\maketitle

Altermagnets are a novel class of collinear antiferromagnets that combine features of both ferromagnets and antiferromagnets: they exhibit nonrelativistic spin splitting in momentum space
and maintain zero net magnetization in real space~\cite{CWu,HYMa,LSmejkal1,LSmejkal2,LDYuan,IIMazin}. 
Magnons in altermagnets are predicted to show analogous splitting between chiral modes~\cite{LSmejkal4,ZLiu,QCui}. 
Despite numerous candidate materials~\cite{HYMa,LSmejkal1,LSmejkal2,LSmejkal3,LDYuan,IIMazin,BJiang,FZhang}, conflicting reports on the existence of altermagnetism persist--both in metallic~\cite{JiayuLiu,YFang,RYChu} and insulating systems~\cite{QFaure,VCMorano}--calling for definitive experimental signatures.
Angle-resolved photoemission spectroscopy (ARPES)~\cite{BJiang,FZhang,JiayuLiu} and neutron scattering~\cite{ZLiu,QFaure,VCMorano} provide direct momentum‑space probes, but transport measurements offer complementary real‑space information.
Electrical Hall measurements are applicable to altermagnetic metals with free carriers~\cite{LSmejkal3,RYChu,YFang}, yet fail in semiconductors and insulators~\cite{IIMazin,HYMa,QCui,QFaure,VCMorano}.

Thermal transport provides a more versatile probe because a temperature gradient exerts statistical forces on a broad range of (quasi)particles, including electrons, magnons, phonons, and their hybrid--even in insulators. 
In the presence of Berry curvature, a thermal gradient induces a transverse heat current, known as the thermal Hall effect (THE).
Linear THE driven by Berry curvature is well established~\cite{RMatsumoto,TQin,RMatsumoto1} and has been widely applied~\cite{LZhang,XZhang,HZhang,RMatsumoto2,TIdeue,CXu,XTZhang,XLi,TSaito,VCvetkovic,XLi2,CXu2}.
Recent theoretical works~\cite{HVarshney,JCLi,YWang} have explored second‑order nonlinear thermal Hall responses in magnonic systems by treating the temperature gradient as an effective electric field $\mathbf{E}=\frac{\bm\nabla T}{T}$, in direct analogy to nonlinear Hall effect of electrons~\cite{Kondo,YGao,ISodemann,ZZDu}.
However, this approach overlooks the spatial dependence of the distribution function and density matrix induced by a temperature gradient, rendering it conceptually incomplete. A proper theory of nonlinear THE for (quasi)particle--including symmetry constraints and practical application--remains lacking.

In this Letter, we investigate the thermal response of altermagnets driven by  Berry curvatures of electrons and magnons. A theoretical framework based on Boltzmann transport theory with relaxation time approximation (RTA) is developed to describe the transverse energy current up to the third-order power of the temperature gradient, applicable to electrons, phonons, magnons and also their hybrid excitation. Based on magnetic point group analysis, we enumerate the symmetry constraints that support dominating third-order response while the linear and second ones vanish, satisfied by planar $d$-wave, $i$-wave and bulk $i$-wave altermagnets with out-of-plane spin orientation, which could be the fingerprints to confirm the altermagnetism.
We calculate the Berry curvature and thermal Hall response of $d$-wave metal KV$_{2}$Se$_{2}$O~\cite{BJiang} and insulator  MnF$_{2}$ to confirm the validity of theoretical prediction and symmetry analysis. 
In KV$_{2}$Se$_{2}$O, both the electrons and magnons
contribute the third-order nonlinear THE when considering spin-orbit coupling and the magnetic dipole-dipole interactions (DDIs). With the spin-spin interactions parameters given by a very recent experiment for MnF$_{2}$~\cite{QFaure}, the magnons alone yield a third-order nonlinear THE.
Thermal Hall conductivity exhibits a $\pi$-period dependence on the direction of temperature gradient, strongly different from the linear and second-order ones. The in-plane spin orientation in these altermagnets
supports linear THE. Based on symmetry analysis, $g$-wave altermagnets with in-plane (out-of-plane) spin order support linear (zero net)response.
Our results provide key transport characteristics for a large amount of altermagnets, even for semiconductors and insulators, helping us distinguish the orientation of N\'{e}el vector.

We start by assuming an $a$-direction temperature gradient $T_{\mathbf{r}}=T+x_{a}(\partial_{a}T)$
and sketching the derivation of transverse thermal Hall current along $b$-direction in the power series of $\partial_{a}T$ up to the third order, which takes the form $J_{Q}^{b}=\kappa_{ba}^{1}(\partial_{a}T)+\kappa_{ba}^{2}(\partial_{a}T)^{2}+\kappa_{ba}^{3}(\partial_{a}T)^{3}$.
The Berry phase effect on the (quasi)particle wavepacket results in the transverse thermal current~\cite{RMatsumoto,LZhang,TQin,RMatsumoto1},
\begin{equation}\label{eq1}
  J_{Q}^{b}=\frac{\varepsilon_{bac}}{\hbar V}\sum_{n\mathbf{k}}\Omega_{n\mathbf{k}}^{c}\int_{E_{n\mathbf{k}}}^{\infty}(E-\mu)\partial_{a}\rho(E,\mathbf{r}) dE,
\end{equation}
where $\varepsilon_{bac}$ is the Levi-Civita tensor, $\mu$ is the chemical potential. $\mu=0$ applies to bosons.
$\mathbf{e}_{a}$, $\mathbf{e}_{b}$, $\mathbf{e}_{c}$ perpendicular to each other and span a
Cartesian coordinate system. $\Omega_{n\mathbf{k}}^{c}$ is the Berry curvature component projected on the $c$-direction.
$\rho(E,\mathbf{r})$ is the particle distribution function.
The above formalism is proposed in electronic~\cite{TQin} and bosonic~\cite{RMatsumoto} systems separately,
and summarized~\cite{LZhang} subsequently to be a general relation for electrons, magnons, phonons and their hybrid excitations.

We adopt the semiclassical theory of wavepacket and
start with the Boltzmann equation in the RTA,
\begin{equation}\label{eq2}
  \partial_{t}\rho+\mathbf{v}_{n\mathbf{k}}\cdot\nabla\rho=-\frac{\rho-\rho^{0}}{\tau_{n\mathbf{k}}},
\end{equation}
the group velocity $\mathbf{v}_{n\mathbf{k}}=\frac{1}{\hbar}\frac{\partial E_{n\mathbf{k}}}{\partial \mathbf{k}}$ and
$\rho^{0}$ is the particle distribution function at equilibrium with local temperature distribution $T_{\mathbf{r}}$. $\tau_{n\mathbf{k}}$ is the relaxation time.
In the steady state, $\partial_{t}\rho=0$, the solution of the distribution function is $\rho=(1+\tau_{n\mathbf{k}}\mathbf{v}_{n\mathbf{k}}\cdot\nabla)^{-1}\rho^{0}=\sum_{l}\rho^{l}$.
The $l$-th order perturbation solution for $\rho$ is
$\rho^{l}=(-\tau_{n\mathbf{k}}\mathbf{v}_{n\mathbf{k}}\cdot\nabla)^{l}\rho^{0}$.
Inserting the solution of $\rho$ into Eq.~\eqref{eq1}, integrating the integral and equating all the terms
in same order of $\partial_{a}T$ yields the thermal current in the formalism $ J_{Q}^{b}=\sum_{l}\kappa_{ba}^{l}(\partial_{a}T)^{l}$.
The first-order thermal Hall conductivity $\kappa_{ba}^{1}=\frac{k_{B}^{2}T}{\hbar V}\sum_{n\mathbf{k}}\varepsilon_{bac}\Omega_{n\mathbf{k}}^{c}c_{2}^{\eta}(\rho_{n\mathbf{k}}^{0}(T))$
is already obtained in the previous work~\cite{TQin,RMatsumoto}. The contributions from second- and third-order THEs are characterized by
\begin{equation}\label{eq3}
  \kappa_{ba}^{2}=-\frac{k_{B}^{2}}{\hbar V}\sum_{n\mathbf{k}}\varepsilon_{bac}\tau_{n\mathbf{k}} v_{n\mathbf{k}}^{a}\Omega_{n\mathbf{k}}^{c}\frac{\partial}{\partial T}[Tc_{2}^{\eta}(\rho_{n\mathbf{k}}^{0}(T))],
\end{equation}
\begin{equation}\label{eq4}
  \kappa_{ba}^{3}=\frac{k_{B}^{2}}{\hbar V}\sum_{n\mathbf{k}}\varepsilon_{bac}(\tau_{n\mathbf{k}} v_{n\mathbf{k}}^{a})^{2}\Omega_{n\mathbf{k}}^{c}\frac{\partial^{2}}{\partial T^{2}}[Tc_{2}^{\eta}(\rho_{n\mathbf{k}}^{0}(T))],
\end{equation}
respectively. $\eta=\pm$. $\eta=+$ applies to fermions with $c_{2}^{+}(x)=\frac{\pi^{2}}{3}+(x-1)(\ln\frac{1-x}{x})^2-2\ln x\ln\frac{1-x}{x}+2\mathrm{Li}_{2}(\frac{x-1}{x})$, and $\rho_{n\mathbf{k}}^{0}(T)$ is the Fermi-Dirac distribution function.
$\eta=-$ applies to bosons with  $c_{2}^{-}(x)=(1+x)(\ln\frac{1+x}{x})^{2}-(\ln x)^{2}-2\mathrm{Li}_{2}(-x)$ and $\rho_{n\mathbf{k}}^{0}(T)$  is the Bose-Einstein distribution.
$\mathrm{Li}_{2}(x)$ is the polylogarithm function. Limitations $c_{2}^{+}(\rho\rightarrow 0)=0$, $c_{2}^{+}(\rho\rightarrow 1)=\frac{\pi^{2}}{3}$, $c_{2}^{-}(\rho\rightarrow 0)=0$, 
$c_{2}^{-}(\rho\rightarrow\infty)=\frac{\pi^{2}}{3}$ indicate that only finite occupation contributes. Details of derivations are presented in the Supplementary Materials (SM)~\cite{SM}. 

The nonlinear thermal Hall response arises from deviations of the distribution function from the equilibrium. As a result, 
$\kappa_{ba}^{2}$ and $\kappa_{ba}^{3}$ depend on the relaxation time. Since the group velocity $\mathbf{v}_{n\mathbf{k}}$ is odd-parity in the $\mathbf{k}$-space,
a dominating $\kappa_{ba}^{2}$ requires the odd-parity distribution of Berry curvature components $\Omega_{n\mathbf{k}}^{c}$ along certain polar axis, 
requiring the inversion symmetry breaking. To obtain a dominant third-order response, even-parity of $\Omega_{n\mathbf{k}}^{c}$
is required to make $\kappa_{ba}^{2}$ vanish. 
The group velocity square in $\kappa_{ba}^{3}$ indicates that the third-order thermal Hall effect is universally intertwined with the linear response. A finite  $\kappa_{ba}^{1}$ implies a nonzero $\kappa_{ba}^{3}$. 
To rule out linear THE, relation $\sum_{\mathbf{k}}\Omega_{n\mathbf{k}}^{c}\delta(E-E_{n\mathbf{k}})=0$ should be satisfied for participated bands. The two constraints greatly reduce the range of symmetry, making us mainly focusing on the black and white magnetic point group, which breaks the time reversal but allows the cooperative transformation of lattice symmetry and time reversal.  
We here conclude that the materials invariant under the following magnetic point groups: 
$mmm$, $m^{\prime}m^{\prime}m$, $m^{\prime}m^{\prime}2$, $2m^{\prime}m^{\prime}$, 
$4^{\prime}$, $\bar{4}^{\prime}$, $4^{\prime}/m$, $4^{\prime}mm^{\prime}$,
$4^{\prime}/mmm^{\prime}$, $6m^{\prime}m^{\prime}$, 
$6/mm^{\prime}m^{\prime}$, $m\bar{3}m^{\prime}$ are expected to give dominating third-order nonlinear THE.
Note that symmetry constraints do not coincide with that of third-order nonlinear electric Hall effect even when Berry curvature dominates~\cite{RYChu,ChPZhang}. 
An analysis when considering the scatterings details instead of RTA also supports above symmetry analysis. The RTA provides 
direct formalism for comprehension. We presented details in SM~\cite{SM}. 
Since the above magnetic point groups describe the symmetries of magnetic materials, not only the electrons but also the spin wave (magnon) in the satisfied materials contribute to the THE.

\begin{figure}[t]
	\centering
	\includegraphics[width=0.48\textwidth]{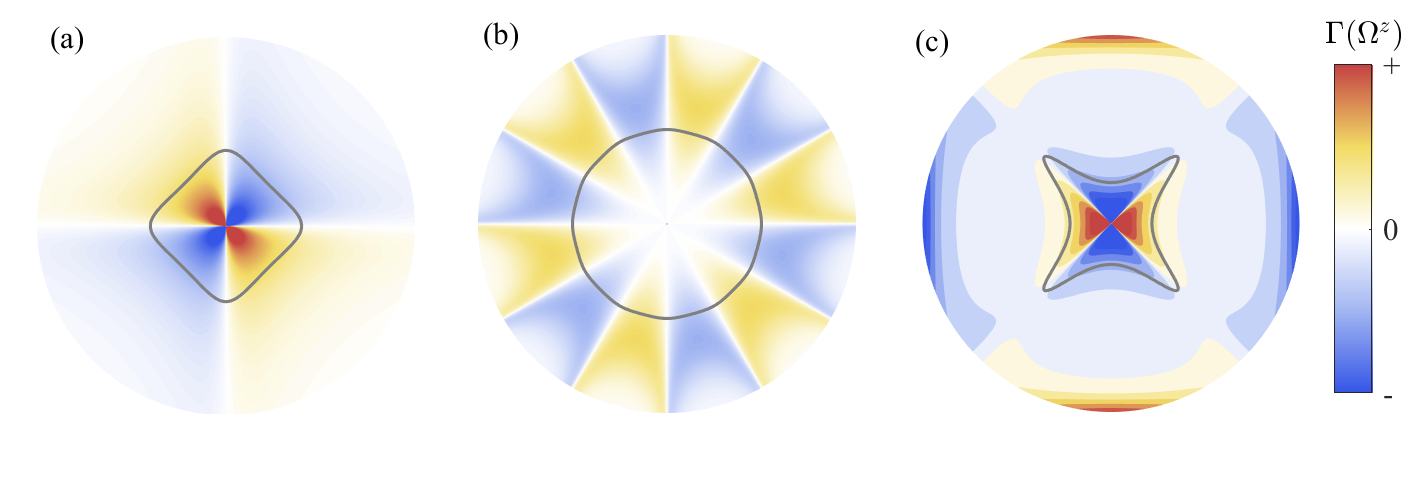}
	\caption{The Berry curvature pattern in the $k_{x}-k_{y}$ plane for the planar $d$-wave (a),
	$i$-wave (b) and bulk $i$-wave (c) altermagnets with the effective Hamiltonian given by Eq.~\eqref{eq5}.
	$t_{J}=0.5$ and $\lambda=0.1$. $\Gamma(\Omega^{z})=\mathrm{sgn}(\Omega^{z})\log_{10}(1+|\Omega^{z}|)$.
	The gray lines denote an constant energy surface. In (c), $k_{z}=1.0$.}\label{fig1}
\end{figure}

Specifically, the planar $d$-wave ($4^{\prime}/m$, $4^{\prime}mm^{\prime}$, $4^{\prime}/mmm^{\prime}$), 
$i$-wave ($6m^{\prime}m^{\prime}$, $6/mm^{\prime}m^{\prime}$) and bulk $i$-wave ($m\bar{3}m^{\prime}$) altermagnets~\cite{LSmejkal2} satisfy the symmetry constraints.
These magnetic point groups make constraints on the spin orientation on the satisfied altermagnets.
For example, the antiferromagnetic order in planar $d$-wave and $i$-wave altermagnets is along the 
$c$-axis ($z$-direction) while the alternating spin splittings are in the perpendicular ($xy$) plane.
We firstly consider the contributions from electrons in altermagnetic metals and narrow-gap semiconductors. 
The SOC should be considered for electronic bands to get finite Berry curvature of electrons.
We apply a unified effective $\mathbf{k}\cdot\mathbf{p}$ electronic Hamiltonian~\cite{LSmejkal2} near the $\Gamma$-point and consider a Rashba-type SOC for the above three 
type altermagnets, 
\begin{equation}\label{eq5}
  H=H_{alt}+\lambda(k_{x}\sigma_{y}-k_{y}\sigma_{x}),
\end{equation}
where $H_{alt}=t_{J}k_{x}k_{y}\sigma_{z}$, $H_{alt}=t_{J}k_{x}k_{y}(3k_{x}^{2}-k_{y}^2)(3k_{y}^{2}-k_{x}^2)\sigma_{z}$ and  
$H_{alt}=t_{J}(k_{x}^{2}-k_{y}^{2})(k_{y}^{2}-k_{z}^{2})(k_{z}^{2}-k_{x}^{2})\sigma_{z}$ correspond to 
planar $d$-wave, $i$-wave and bulk $i$-wave, respectively. 
The Berry curvature distribution of $\Omega_{\mathbf{k}}^{z}$ in the $k_{x}$-$k_{y}$ plane is shown in Fig.~\ref{fig1}, directly indicating 
the vanishing $\kappa_{ba}^{1}$ and $\kappa_{ba}^{2}$.

\begin{figure}[t]
  \centering
  \includegraphics[width=0.48\textwidth]{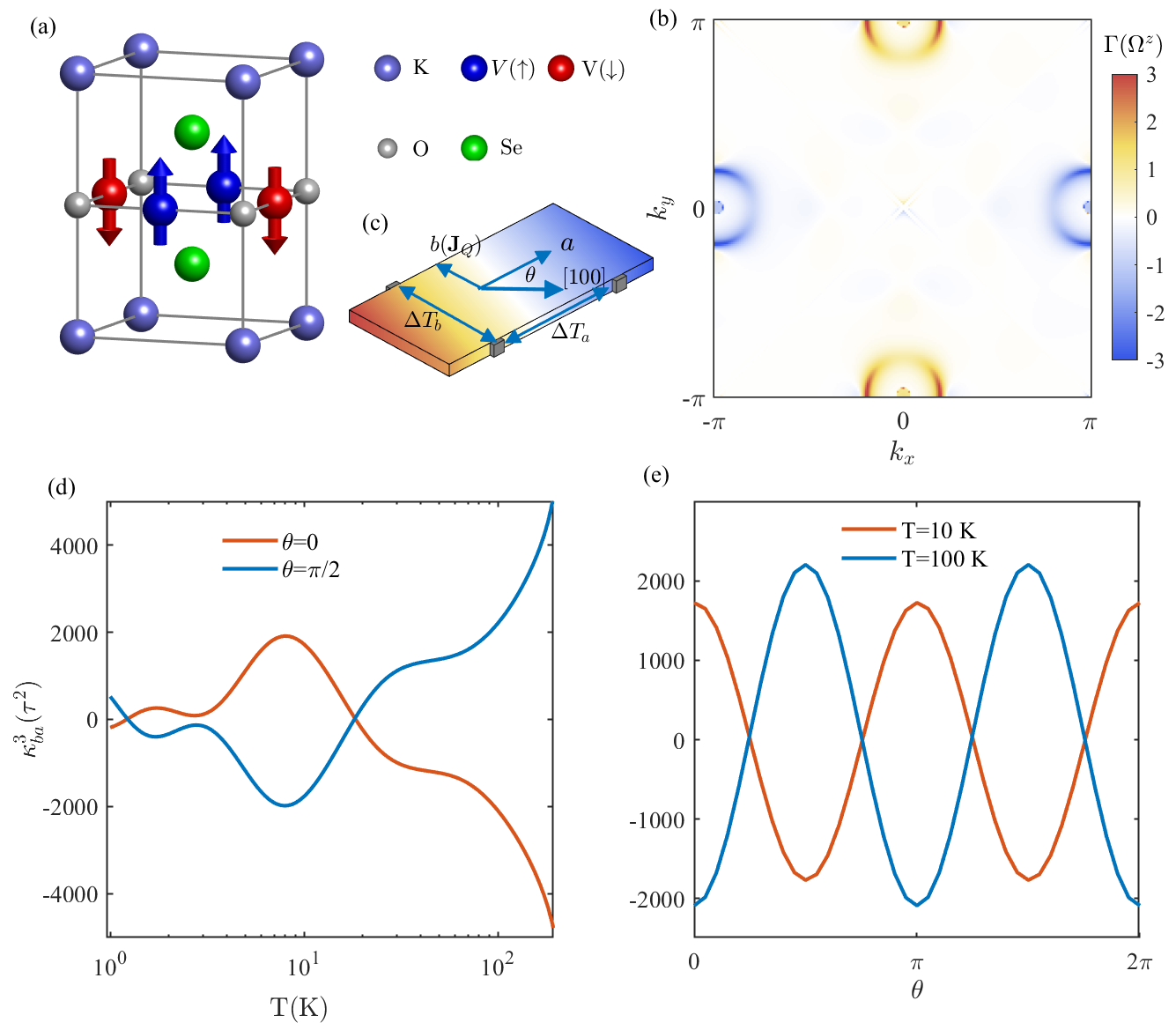}
  \caption{The THE of electrons in altermagnetic metal KV$_{2}$Se$_{2}$O. 
  (a) The crystal structure of KV$_{2}$Se$_{2}$O. (b) The distribution of $z$ component electronic Berry curvature with log scale $\Gamma(\Omega^{z})=\mathrm{sgn}(\Omega^{z})\log_{10}(1+|\Omega^{z}|)$
  in the $k_{z}=0$ plane. The unit of Berry curvature is \AA$^{2}$.  (c) Illustration of a thermal Hall detection setup with the temperature gradient in the $xy$ plane. (d) $\kappa_{ba}^{3}$ with respect to the temperature at two angles.
  (e) $\kappa_{ba}^{3}$ with respect to the angle of temperature gradient at two different temperatures.}\label{fig2}
\end{figure}

We adopt the realistic materials to apply our theory for detailed discussions.  
The recent experiment reported a $d$-wave altermagnetic metal KV$_{2}$Se$_{2}$O~\cite{BJiang},  
with the crystal structure and spin order shown in Fig.~\ref{fig2} (a), invariant under the magnetic point group 
$4^{\prime}/mmm^{\prime}$. We employ the state-of-the-art first-principles calculations 
to compute the electronic bands including the SOC and the corresponding Berry curvature in the momentum space.  
Fermi energy cuts through two bands and others are far separated~\cite{BJiang}. 
The The distribution of $z$-component Berry curvature in the Brillouin zone for one of the two bands with higher energy 
is plotted in Fig.~\ref{fig2} (b), which satisfies the two constraints. 
$x$ and $y$ components of Berry curvature vanish.
The summation of $\Omega^{z}$ show even parity in the $k_{x}-k_{y}$ plane and along $k_{z}$ direction. 
The Berry curvature of the other band with lower energy hold the same features. 
To investigate the thermal Hall response, a temperature gradient within the $xy$ plane at an angle $\theta$ respect to the $x$ axis (or [100] crystal direction) is applied, as illustrated in Fig.~\ref{fig2} (c). 
The Berry curvature generates a nonzero energy current within the $xy$ plane and perpendicular to the temperature gradient. 
Third-order nonlinear thermal Hall response dominates, giving a finite $\kappa_{ba}^{3}$.
Since the Berry curvature locates in a very narrow region in the momentum space,
we can adopt the a single relaxation time for the calculations as an appropriate approximation. 
In Fig.~\ref{fig2} (d), we plotted the temperature dependent $\kappa_{ba}^{3}$ at two $\theta$ values. 
$\kappa_{ba}^{3}$ becomes larger at high temperatures, convenient for the experimental detections.
From Fig.~\ref{fig2} (e), $\kappa_{ba}^{3}$ is strongly dependent on the direction of temperature gradient, exhibiting 
a $\pi$-period, absent in the first and second-order thermal Hall response.
This $\pi$-period is a distinguished feature of the third-order thermal Hall response due to 
$(v_{n\mathbf{k}}^{a})^{2}$ term in Eq.~\ref{eq4}. Since $(v_{n\mathbf{k}}^{a})^{2}=(v_{n\mathbf{k}}^{x})^{2}\cos^{2}\theta+(v_{n\mathbf{k}}^{y})^{2}\sin^{2}\theta+2v_{n\mathbf{k}}^{x}v_{n\mathbf{k}}^{y}\sin2\theta$, we can write 
$\kappa_{ba}^{3}(\theta)=\kappa_{ba}^{3,xx}\cos^{2}\theta+\kappa_{ba}^{3,yy}\sin^{2}\theta+\kappa_{ba}^{3,xy}\sin2\theta$. The relation $\kappa_{ba}^{3}(\theta+\pi)=\kappa_{ba}^{3}(\theta)$ is naturally satisfied. 
Besides KV$_{2}$Se$_{2}$O, the other $d$-wave altermagnetic metals FeSb$_{2}$~\cite{LSmejkal1,LSmejkal2},
Rb$_{1-\delta}$V$_{2}$Te$_{2}$O~\cite{FZhang} are also candidates.
By matching the above symmetries and taking the SOC into account in the calculations,
it is probable to discover many other materials.

We turn to the magnon contributed thermal Hall response, which emerges not only in altermagnetic metals 
but also in semiconductors and insulators. 
Several experimental studies have confirmed the chiral splitting behavior of the magnons arising from
the altermagnetic exchange interactions between local spins~\cite{LSmejkal4,ZLiu,QCui}--a phenomenon analogous to that observed in electronic systems.
A general spin-spin interaction Hamiltonian in altermagnets is given by
\begin{equation}\label{eq6}
  H=\sum_{i<j}(J_{ij}\delta_{\alpha\beta}+K_{ij}^{\alpha\beta})S_{i}^{\alpha}S_{j}^{\beta}-D\sum_{i}(S_{i}^{z})^{2},
\end{equation}
where $J_{ij}$ represents the exchange interactions. The DDIs between local spins are described by
$K_{ij}^{\alpha\beta}=g^{2}\mu_{B}^{2}(r_{ij}^{2}\delta_{\alpha\beta}-3r_{ij}^{\alpha}r_{ij}^{\beta})/2r_{ij}^{5}$,
where $\mathbf{r}_{ij}=\mathbf{r}_{i}-\mathbf{r}_\mathbf{j}$ being the vector connecting two spins.
The last term accounts for magnetic anisotropy. It is worth noting that other complex interactions are not necessary to capture the phenomenon of interest here. 

\begin{figure}[t]
	\centering
	\includegraphics[width=0.48\textwidth]{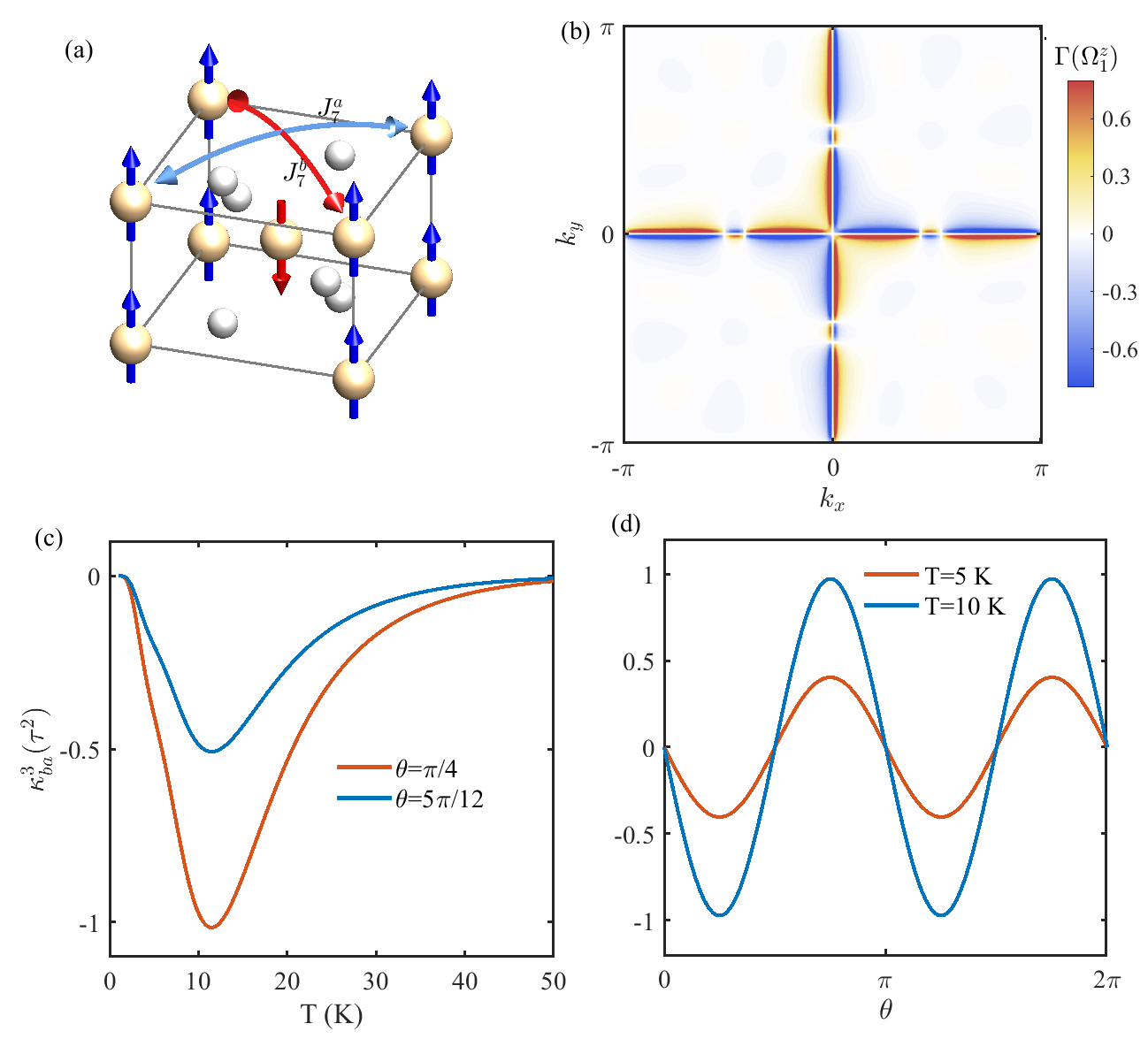}
	\caption{The THE of magnons in altermagnetic insulator MnF$_{2}$.
		(a) The crystal structure of MnF$_{2}$. (b) The distribution of $z$ component magnon Berry curvature with log scale
		$\Gamma(\Omega_{1}^{z})=\mathrm{sgn}(\Omega_{1}^{z})\log_{10}(1+|\Omega_{1}^{z}|)$
		in the $k_{z}=0$ plane. The unit of Berry curvature is \AA$^{2}$. (c) $\kappa_{ba}^{3}$ with respect to the temperature at two angles.
		(d) $\kappa_{ba}^{3}$ with respect to the angle of temperature gradient at two different temperatures.
	}\label{fig3}
\end{figure}

We apply the Holstein-Primakoff transformations to get the magnon spectrum~\cite{SM} and magnon Berry curvature. 
Although we do not have detailed parameters of the spin-spin interactions in KV$_{2}$Se$_{2}$O and above other candidates, 
recent experiment revealed the $d$-wave altermagnetic configurations of spin-spin interactions
in insulator MnF$_{2}$~\cite{QFaure}, with the structure and spin order depicted in Fig.~\ref{fig3} (a), invariant 
under the magnetic point group $4^{\prime}/mmm^{\prime}$. 
In MnF$_{2}$, the ground state spin configuration is along the $c$-axis due to the DDI and $D=0$,
The exchange parameters are $J_{1}=-0.075$ meV, $J_{2}=0.287$ meV, $J_{3}=-0.012$ meV for the three nearest-neighbor exchange interactions~\cite{QFaure}.
$g=1.0$ and the lattice constant is $a=b=4.873$ \AA, $c=3.311$ \AA.
The anisotropic exchange interactions along the in-plane diagonal directions are given by
$J_{7}^{a}=-0.006$ meV and $J_{7}^{b}=-0.002$ meV~\cite{QFaure}, as illustrated in Fig.~\ref{fig3} (a).
The combination of DDIs and $J_{7}^{a}\neq J_{7}^{b}$ splits the magnon bands and generates
finite Berry curvature with the parity in momentum space summarized in Table~\ref{table1}.
All components of magnon Berry curvature exhibit even parity in $\mathbf{k}$ space and the summations over constant energy surfaces vanish, leading to vanishing 
THE of magnons based on the previous theory. 

Under the temperature gradient shown in  Fig.~\ref{fig2} (b),
$z$-component of magnon Berry curvature, $\Omega_{n}^{z}$,
contributes a nonzero net $\kappa_{ba}^{3}$ and the transverse energy current is along the $b$-direction, same to the electronic contribution. 
This is because a nonzero summation over $\mathbf{k}$-space for $\kappa_{ba}^{3}$ in Eq.~\eqref{eq4}
requires even parity of Berry curvature component in the $k_{x}$-$k_{y}$ plane and $k_{z}$ direction due to the group velocity square 
[Fig.~\ref{fig3} (b) and Table~\ref{table1}]. The contributions from $\Omega_{n}^{x}$ and $\Omega_{n}^{y}$ vanish. 
Due to the low N\'{e}el temperature and the narrow distributed Berry curvature in momentum space, low-energy magnon states dominates and we adopt single relaxation time to perform the calculations. 
Fig.~\ref{fig3} (c) shows the dependence of $\kappa_{ba}^{3}$ with respect to the temperature
at $\theta=45^{\circ}$ and $75^{\circ}$, respectively.
The direction change of the temperature gradient modify the value of $\kappa_{ba}^{3}$ dramatically, giving a same $\pi$-period dependence [Fig.~\ref{fig3} (d)], which is
a distinct feature of the third-order nonlinear THE.
Parity of $\Omega_{n}^{x}$ ($\Omega_{n}^{y}$) shares the similar behavior of $\Omega_{n}^{z}$, as shown in Table~\ref{table1}.
A temperature gradient in the $yz$ ($xz$) plane will also give dominant third-order THE contributed by
$\Omega_{n}^{x}$ ($\Omega_{n}^{y}$).
As a result, temperature gradient along any direction in MnF$_{2}$ will generate the nonlinear third-order THE. 
Above altermagnetic metals are the candidates to exhibit THEs contributed by magnons. 
In real experiments, the detection of THEs should consider the both contributions.  
Insulators MnF$_{2}$, FeF$_{2}$, MnO$_{2}$~\cite{LSmejkal1,LSmejkal2}, V$_{2}$Se$_{2}$O~\cite{HYMa}, Cr$_{2}$Te(Se)$_{2}$O~\cite{QCui} etc. show the THE contributed only by magnon.

\begin{table}[t]
  \centering
  \caption{The parity of the magnon Berry curvature along three directions for both out-of-plane and in-plane N\'{e}el vectors.
  $+$ ($-$) denotes even (odd) parity.}\label{table1}
  \setlength{\tabcolsep}{8pt}
  \renewcommand{\arraystretch}{1.5}
  \begin{tabular}{c|ccc|ccc}
  \hline
   N\'{e}el vector& \multicolumn{3}{c|}{Out-of-plane} &  \multicolumn{3}{c}{In-plane (x)} \\
  \hline
   Berry curvature & $\Omega_{n}^{x}$ & $\Omega_{n}^{y}$ & $\Omega_{n}^{z}$ & $\Omega_{n}^{x}$ & $\Omega_{n}^{y}$ & $\Omega_{n}^{z}$  \\
   \hline
   $k_{x}$ & $+$ & $-$  & $-$ & $-$ & $+$ & $+$  \\
   \hline
   $k_{y}$ & $-$ & $+$ & $-$ & $-$ & $+$ & $-$    \\
   \hline
   $k_{z}$ & $-$ & $-$ & $+$ & $+$ & $+$ & $-$ \\
  \hline
  \end{tabular}
\end{table}

The thermal Hall response depends on the orientation of N\'{e}el vector in altermagnets. 
For a negative $D<D_{c}<0$, the easy-plane anisotropy makes the N\'{e}el vector along in-plane direction.
$D=-0.05$ meV is assumed for the calculations.
When N\'{e}el vector lies in-plane, for instance, along the $x$-direction. The magnetic point group describing this spin 
configuration is $2^{\prime}/m^{\prime}m^{\prime}m$, not in the above list for dominating third-order THE and 
supporting linear THE. 
The parity of magnon Berry curvature is different in this configuration, as summarized in Table~\ref{table1}.
The even parity of $\Omega_{n}^{y}$ in any direction indicates the dominating linear thermal Hall response.
$\kappa_{ba}^{1}$ will always be present due to nonzero $\Omega_{n}^{y}$. A recent 
work~\cite{RHoyer} also discussed linear THE in MnF$_{2}$ when the  N\'{e}el vector is not out-of-plane.
For the electron contribution in altermagnetic metals, 
it is no coincidence that a recent theoretical work reported electron contributed linear THE in RuO$_{2}$~\cite{XZhou}, same crystal structure to MnF$_{2}$ and the N\'{e}el orientation
is in-plane with the same magnetic symmetry.  
This conclusion is also valid for the $i$-wave altermagnets. 
For $g$-wave altermagnets, an out-of-plane N\'{e}el vector ($4/mm^{\prime}m^{\prime}$ for planar one and $6^{\prime}/m^{\prime}mm^{\prime}$ for bulk one) implies zero net THE while in-plane 
one indicates linear THE, based on the symmetry analysis. 
In the absence of altermagnetic splittings, THE vanishes. For example, 
in the absence of altermagnetic configuration in the spin-spin interactions, i.e., when $J_{7}^{a}=J_{7}^{b}$ in 
MnF$_{2}$~\cite{QFaure,VCMorano}, the Berry curvature vanishes, indicating zero THE~\cite{SM}. 
Although DDI is small, the third-order thermal Hall effect exists in a wide range of $J_{7}^{a}-J_{7}^{b}$~\cite{SM} even when the energy scale of splitting is far larger than DDI. 
Above results and conclusions are the key transport characteristics in altermagnets. 
Detecting the thermal response in the candidate materials help us to confirm whether altermagnetism exists 
and also identify the orientation of N\'{e}el vector.

Here we discuss the experimental detection. To date, numerous works have observed linear THE in nonmagnetic materials, magnets, and superconductors~\cite{HZhang,CXu,CXu2,VCvetkovic,XLi,XLi2}. 
Experiments usually adopt three-terminal structures to detect temperature differences, as illustrated in Fig.~\ref{fig2} (c). The longitudinal and transverse temperature differences are 
$\Delta T_{a}$ and $\Delta T_{b}$, respectively. When third-order THE dominates, $J_{Q}^{b}=\kappa\frac{\Delta T_{b}}{L_{b}}=\kappa_{ba}^{3}(\frac{\Delta T_{a}}{d})^{3}$, where $L_{b}$ ($d$) is the distance between two transverse (longitudinal) thermocouple devices, and $\kappa$ is the thermal conductivity of the material. Above relation indicates
$\Delta T_{b}=\frac{\kappa_{ba}^{3}L_{b}}{\kappa {d}^{3}}(\Delta T_{a})^{3}$. Since $\Delta T_{a}$ is tunable by heater power, the observation of $\Delta T_{b}\propto(\Delta T_{a})^{3}$
identifies the dominance of third-order response. Based on previous experiments, we estimate the parameter range for
the observable magnitude of signals.
For magnons, recent experiments in insulating magnets~\cite{HZhang}
indicate a temperature gradient about $(\partial_{a}T)\sim10^{3}-10^{4}$ K/m.
An estimated relaxation time $\tau=10^{-6}\sim10^{-5}$ s for low energy and small momentum magnons~\cite{KeWang} makes the signal $\kappa_{ba}^{3}\times(\partial_{a}T)^{2}\sim 10^{-4}-10^{-2}$ WK$^{-1}$m$^{-1}$ in MnF$_{2}$ comparable to 
the first-order thermal Hall conductivity $\kappa_{ba}^{1}$ in previous works~\cite{HZhang}.
For electrons, a relaxation time $\tau=10^{-6}$ s and temperature gradient about $(\partial_{a}T)\sim10^{3}$ K/m make
$\kappa_{ba}^{3}\times(\partial_{a}T)^{2}$ comparable to $\kappa_{ba}^{1}$. 
The third-order thermal Hall response exhibits distinct dependencies on the relaxation time, temperature, the magnitude, and direction of the thermal gradient, providing clear and distinguishable features for experimental identification. 

In summary, we have developed a theory of nonlinear THE in altermagnets and demonstrated that a third‑order nonlinear THE arises in $d$-wave and other symmetry‑satisfying altermagnets. This nonlinear THE receives contributions not only from electrons in altermagnetic metals but also from magnons in insulators. Our work provides key thermal‑response signatures of altermagnets, with strong potential for confirming altermagnetism in candidate materials.

This work is supported by the NSFC (Grant Nos. 12474050 and 12488101) and
the MOST of China (Grant No. 2022YFA1204700). Y.-M. Li acknowledges helpful discussions with
Prof. Wei-Nan Lin, Long Xu from Xiamen University and Prof. Yongwei Huang from Ningxia University.

\end{document}